\documentclass[lettersize,journal]{IEEEtran}
\usepackage{amsmath,amsfonts,amssymb,bm,amsthm}
\usepackage{color}
\usepackage{algorithmic}
\usepackage{array}
\usepackage{caption}
\usepackage[justification=raggedright,singlelinecheck=false,font=footnotesize]{caption}
\usepackage[subrefformat=parens,labelformat=simple,justification=centerlast]{subcaption}
\usepackage{textcomp}
\usepackage{algorithm, algorithmic}
\usepackage{stfloats}
\usepackage{url}
\usepackage{verbatim}
\usepackage{graphicx}
\def\BibTeX{{\rm B\kern-.05em{\sc i\kern-.025em b}\kern-.08em
    T\kern-.1667em\lower.7ex\hbox{E}\kern-.125emX}}
\usepackage{balance}
\usepackage{threeparttable}
\usepackage{pifont}
\usepackage{bbding}
\usepackage{multirow}
\usepackage{makecell}
\usepackage{booktabs}
\usepackage{epstopdf}
\usepackage{epsfig, latexsym, times}
\usepackage{float}
\usepackage{subcaption}
\usepackage{cases}
\usepackage{stfloats}
\usepackage[colorlinks=true, allcolors=blue]{hyperref}
\usepackage{makecell}
\usepackage{graphicx}
\usepackage{textcomp}
\usepackage{array} 
\usepackage{longtable}
\usepackage{booktabs}
\usepackage{float}

\usepackage{blkarray}
\usepackage{diagbox}

\usepackage{ragged2e}



\renewcommand{\proofname}{Proof}

\newtheoremstyle{italicheader}     
{0pt}                            
{0pt}                            
{\normalfont}                    
{1em}                               
{\itshape}                       
{:}                              
{.5em}                           
{}                               

\theoremstyle{italicheader}

\begin{document}
\title{\LARGE \color{black}Coverage Performance Analysis of FAS-enhanced LoRa Wide Area Networks under both Co-SF and Inter-SF Interference}
\author{Gaoze Mu,~\IEEEmembership{Member,~IEEE,} Yanzhao Hou,~\IEEEmembership{Member,~IEEE,} Mingjie Chen, Yuanyu Hu, Yongan Zheng,\\ Qimei Cui,~\IEEEmembership{Senior Member,~IEEE,} Xiaofeng Tao,~\IEEEmembership{Senior Member,~IEEE}



\thanks{All authors are with Beijing University of Posts and Telecommunications, Beijing 100876, China (e-mail: \{mugz; houyanzhao; chenmingjie; huyuanyu; yonganzheng; cuiqimei; taoxf\}@bupt.edu.cn). (\textit{Corresponding author: Yanzhao Hou.})}

}

\maketitle
\begin{abstract}
	This paper presents an analytical framework for evaluating the coverage performance of the fluid antenna system (FAS)-enhanced LoRa wide-area networks (LoRaWANs). We investigate the effects of large-scale pathloss in LoRaWAN, small-scale fading characterized by FAS, and dense interference  (i.e., packet collisions under the ALOHA protocol) arising from randomly deployed end devices (EDs). Both co-spreading factor (co-SF) interference (with the same SF) and inter-SF interference (with different SFs) are introduced into the network, and their differences in physical characteristics are also considered in the analysis. Additionally, simple yet accurate statistical approximations of the FAS channel envelope and power are derived using the extreme-value theorem. Based on the approximated channel expression, the theoretical coverage probability of the proposed FAS-enhanced LoRaWAN is derived. Numerical results validate our analytical approximations by exhibiting close agreement with the exact correlation model. Notably, it is revealed that a FAS with a normalized aperture of $1\times1$ can greatly enhance network performance, in terms of both ED numbers and coverage range.
\end{abstract}


\section{Introduction}
\IEEEPARstart{R}{ecently}, with the rapid growth of the Internet-of-
Things (IoT), LoRa wide-area networks (LoRaWANs) have gained considerable attention for enabling long-range, low-power, and low-data-rate massive machine-type communications (mMTC) \cite{bibitem1}. Despite these advantages, LoRa suffers severe performance degradation in fading environments \cite{bibitem2}, and its ALOHA-based access protocol poses a critical challenge for interference management \cite{bibitem3}. While multiple-input multiple-output (MIMO) techniques can effectively address both issues, their applicability in LoRa systems is limited by power consumption and hardware complexity. As a low-complexity alternative, the fluid antenna system (FAS) that advocates position and shape flexibility in antennas has emerged as a new degree of freedom. \cite{bibitem4} explored the integration of FAS into LoRa, demonstrating notable improvements in link-level error performance. However, it remains unclear whether such gains persist under severe large-scale pathloss and dense end-device (ED) deployments, where network performance is predominantly affected by the interference and collisions.

From the perspective of the LoRaWAN, the coexistence of co-spreading factor (co-SF) interference (with the same SF) and inter-SF interference (with different SFs) significantly complicates the performance analysis of LoRa networks, as the two exhibit fundamentally different physical characteristics \cite{bibitem5,bibitem6}. Specifically, co-SF interference coherently accumulates at the receiver, producing detection peaks similar to the desired signal. In contrast, inter-SF interference is largely suppressed by the quasi-orthogonality among SFs, spreading its energy across the spectrum and behaving as noise \cite{bibitem3,bibitem7}. Most existing works analyze these two interference types separately \cite{bibitem6}, \cite{bibitem8}, consider only one of them \cite{bibitem9,bibitem10,bibitem11}, or neglect their differences altogether \cite{bibitem12,bibitem13}. These motivate the need for analytical frameworks that explicitly capture the distinct behaviors of both co-SF and inter-SF interference. Additionally, although spatial-domain techniques, such as MIMO, reconfigurable intelligent surfaces (RIS), and FAS, have been investigated at the link level for LoRa systems, as summarized in \cite{bibitem4}, their network-level performance gains in LoRaWAN remain largely unexplored, even in the well-established MIMO scenarios.

Building on the above works, we believe that evaluating the spatial gain of FAS on LoRaWAN is of significant relevance to both sides. In this paper, we aim to evaluate the coverage performance of FAS-enhanced LoRaWAN under both co-SF and inter-SF interference, while explicitly accounting for the distinct physical characteristics of the two interference types. For the FAS, simple yet accurate statistical expressions for both the channel envelope and power are derived. Based on the above, we obtain accurate analytical results for the coverage probability of the proposed network. Numerical results demonstrate that the proposed FAS scheme significantly enhances coverage probability and range compared to the conventional single-antenna LoRa network, in large-scale regions with randomly deployed and activated EDs.

\section{Network Model}\label{sec:model}
\subsection{LoRaWAN Model}
In this paper, we consider a single-cell uplink LoRa network, where a FAS-equipped gateway is located at the center of the 2-dimensional (2D) annulus $\mathcal{A}\subset\mathbb{R}^2$ with radii $R_1$ and $R_2$ ($R_1<R_2$). The traditional single-antenna EDs are randomly distributed inside $\mathcal{A}$ according to a homogeneous Poisson point process (HPPP) $\mathcal{P}_{L}$ with the density of $\lambda_L=\frac{N}{\pi(R_2^2-R_1^2)}$, where $N$ is the average number of EDs. The activity of the transmission of the EDs follows a HPPP $\mathcal{P}^{(m)}_{A}$, with the density of $\lambda^{(m)}_A=\rho^{(m)} p^{(m)} \lambda_L$, where $\rho^{(m)}$ and $p^{(m)}$ are the ED duty-cycle and LoRa SF allocation probability for $SF=m$, where $m\in\{7,8,...,12\}$. Then, for a given transmission interval, the average number of active EDs with $SF=m$ is given by
\begin{equation}\label{avg_N}
	\overline{N}^{(m)}=\rho^{(m)} p^{(m)} N.
\end{equation}
The duty-cycle $\rho^{(m)}$ of the EDs can be represented as $\rho^{(m)}=\frac{L_{\mathrm{pac}}^{(m)}}{T_{\mathrm{in}}G^{(m)}}$, where $T_\mathrm{in}$ is the average packet arrival time. Besides, $G^{(m)}=\frac{mBW}{2^m}\left(\frac{4}{4+{CR}}\right)$ is the bit rate of SF $m$, and ${CR}\in\{1,2,...,4\}$ is the coding rate, corresponding to a cyclic coding rate of $\left\{\frac{4}{5},\frac{4}{6},...,\frac{4}{8}\right\}$. $BW$ is the transmission bandwidth, with a typical value of $\{50,125,500\}$~kHz. Moreover, $L_{\mathrm{pac}}^{(m)}$ is the packet length, which is given by \cite{bibitem9}
\begin{equation}\label{pac_len}
	L_{\mathrm{pac}}^{(m)}=m\!\left(
	\!\!\begin{array}{lc}
		\left(N_\mathrm{pre}+4.25\right)+8+(4+{CR}) \\
		\times\max\left\{\left\lceil \frac{8{L}_\mathrm{pl}-4m+28+L_\mathrm{crc}-L_\mathrm{hd}}{4m} \right\rceil,0\right\}\\
	\end{array}\!\!\!\right),
\end{equation}
where $L_\mathrm{pl}$ is the payload length. Moreover, $L_\mathrm{crc}$ and $L_\mathrm{hd}$ indicate the length of the cyclic redundancy check and frame header. $N_\mathrm{pre}$ is the number of preamble symbols. Besides, the SF allocation is consistent with \cite{bibitem9,bibitem14}, employing a fair-collision scheme. Then, the probability of occurrence for $SF=m$ can be expressed as	$p^{(m)}=\frac{m}{2^m}\sum^{12}_{\tilde{m}=7}\frac{\tilde{m}}{2^{\tilde{m}}}$, which guarantees a fair collision probability among all SFs and  enhances the network capacity.

\subsection{Channel Model}
Both large-scale pathloss and small-scale fading are considered in the channel model, which describes the network's and FAS's characteristics, respectively. Specifically, the large-scale fading between the $k$-th EDs and the LoRa gateway under a distance of $r_k$ is modeled as
\begin{equation}\label{h-large}
	PL_k=\left({4\pi f_\mathrm{c}}/{c}\right)^2r_k^\beta=K_0r_k^\beta,
\end{equation}
where $K_0$ and $\beta\geq2$ are the pathloss constant and exponent, respectively.  Moreover, $f_\mathrm{c}$ is the carrier frequency and $c=3\times10^8$ m/s is the speed of electromagnetic waves.

Here, we assume a 2D $L_1\times L_2=L$-ports planar FAS with an aperture of $(W_1\times W_2)\lambda=W\times\lambda$ is equipped on the LoRa gateway, where $\lambda=c/f_\mathrm{c}$ is the wavelength of the carrier. The block-correlation model in \cite{bibitem15} is used to generate the small-scale fading, where the correlation factor between any FAS ports $l=(l_1,l_2)$ and $\tilde{l}=(\tilde{l}_1,\tilde{l}_2)$ is given by Jake's model as
\begin{equation}\label{ecm}
		\mathbf{\Sigma}_{l,\tilde{l}}={J_0}\left(\!\!2\pi\sqrt{\left(\frac{\big|l_1-\tilde{l}_1\big|W_1}{L_1-1}\right)^2\!\!+\left(\frac{\big|l_2-\tilde{l}_2\big|W_2}{L_2-1}\right)^2}\right)\!,
\end{equation}
for $l_1\in\{1,\cdots,L_1\}$ and $l_2\in\{1,\cdots,L_2\}$. Besides, $J_0(\cdot)$ is the $0$-th order Bessel function. 

Based on \cite{bibitem15}, (\ref{ecm}) can be approximated by the block diagonal matrix as  $\tilde{\mathbf{\Sigma}}=\mathrm{Blkdiag}(\mathbf{A}_1,\mathbf{A}_2,\cdots,\mathbf{A}_{A})\in\mathbb{R}^{L\times L}$, where $\mathbf{A}_{a}=\mathrm{Toeplitz}\left(1,\mu_a^2,\mu_a^2,\cdots,\mu_a^2\right)\in\mathbb{R}^{L_a\times L_a}$,
by fitting the eigenvalues of $\mathbf{\Sigma}$ in (\ref{ecm}). Note that $A$ and $L_a$, $a\in\{1,\dots,A\}$ are the number of blocks and the dimensions of each block, respectively. Then, each channel realization between $k$-th ED and $l$-th port at the gateway is given by
\begin{multline}\label{h-channel}
		h_{k,l}\triangleq h_{k,\left(a,b_a\right)}=\sqrt{1-\mu_a^2} x_{k,\left(a,b_a\right)}+\mu_a x_{k,\left(0,b_0\right)}\\
		+j\left(\sqrt{1-\mu_a^2}y_{k,\left(a,b_a\right)}+\mu_a y_{k,\left(0,b_0\right)}\right),
\end{multline}
for $k\in\{1,\dots,K\}$, $l\in\{1,\dots,L\}$ and $b_a\in\{1,\dots,L_a\}$. Besides, $A$, $\mu_a$ and $L_a$ are obtained by the variable block-correlation model (VBCM) \cite[Algorithm 1]{bibitem16}. Each $l$ has a one-to-one relationship with $(a,b_a)$ and we have $L\triangleq\sum_{a=1}^{A}{L_a}$. Moreover, $x_{k,(0,b_0)}$, $y_{k,(0,b_0)}$, $x_{k,(a,b_a)}$ and $y_{k,(a,b_a)}\sim\mathcal{N}(0,0.5)$ are independent and identically distributed (i.i.d.) random variables (RVs).

\subsection{Coverage Probability}
We assume that the FAS at the gateway can switch based on the channel estimation results for the desired signal ($k=0$ in \eqref{h-channel}) obtained from the extended preamble\footnote{In \cite{bibitem17}, the preamble of LoRa can be extended from 6 to 65535 symbols, which significantly exceeds the number of ports.}. Therefore, the FAS-port is chosen as  $\hat{l}=\mathrm{arg}\max\limits_{l\in\{1,\cdots,L\}}\left\{\left|h_{0,1}\right|,\dots,\left|h_{0,L}\right|\right\}$. Then the power of the desired ED with $SF=m$ at the receiver can be expressed as
\begin{equation}\label{S_0}
		{S_0}\!=\max\limits_{l\in\{1,\cdots,L\}}\frac{P_\text{t}}{K_0\big(r_0\big)^{\beta}}\left|h_{0,l}\right|^2
		{\triangleq\frac{P_\varsigma}{\big(r_0\big)^{\beta}}\big|h_{0,\hat{l}}\big|^2},
\end{equation}
where $P_{\mathrm{t}}$ is the baseline transmit power and $P_\varsigma={P_\text{t}}/{K_0}$.
	
\begin{figure}[t]\label{fig:1}
	\centering
	\includegraphics[width=0.49\textwidth]{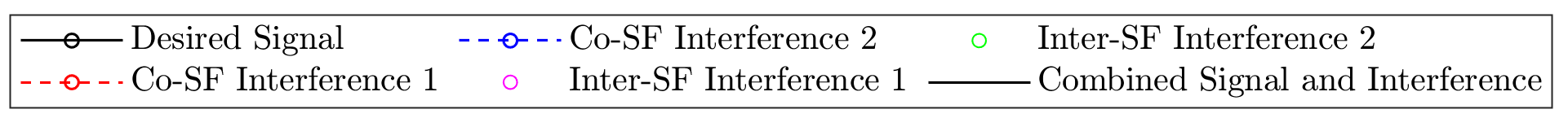}\\[0.5em]
	\begin{subfigure}[b]{0.24\textwidth}
		\centering
		\includegraphics[width=\textwidth]{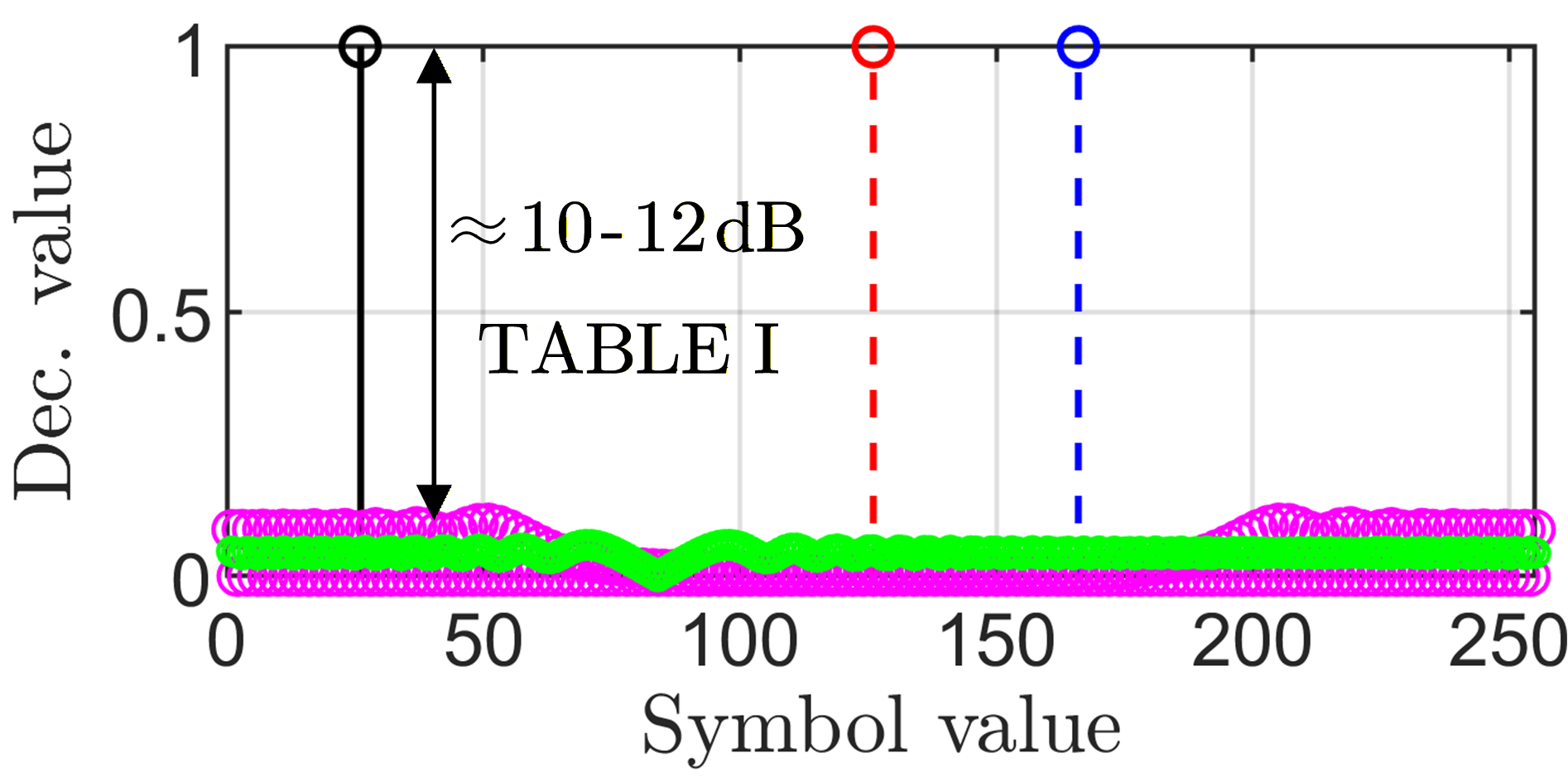}
		\caption{}
		\label{sub_fig:1}
	\end{subfigure}
	\begin{subfigure}[b]{0.24\textwidth}
		\centering
		\includegraphics[width=\textwidth]{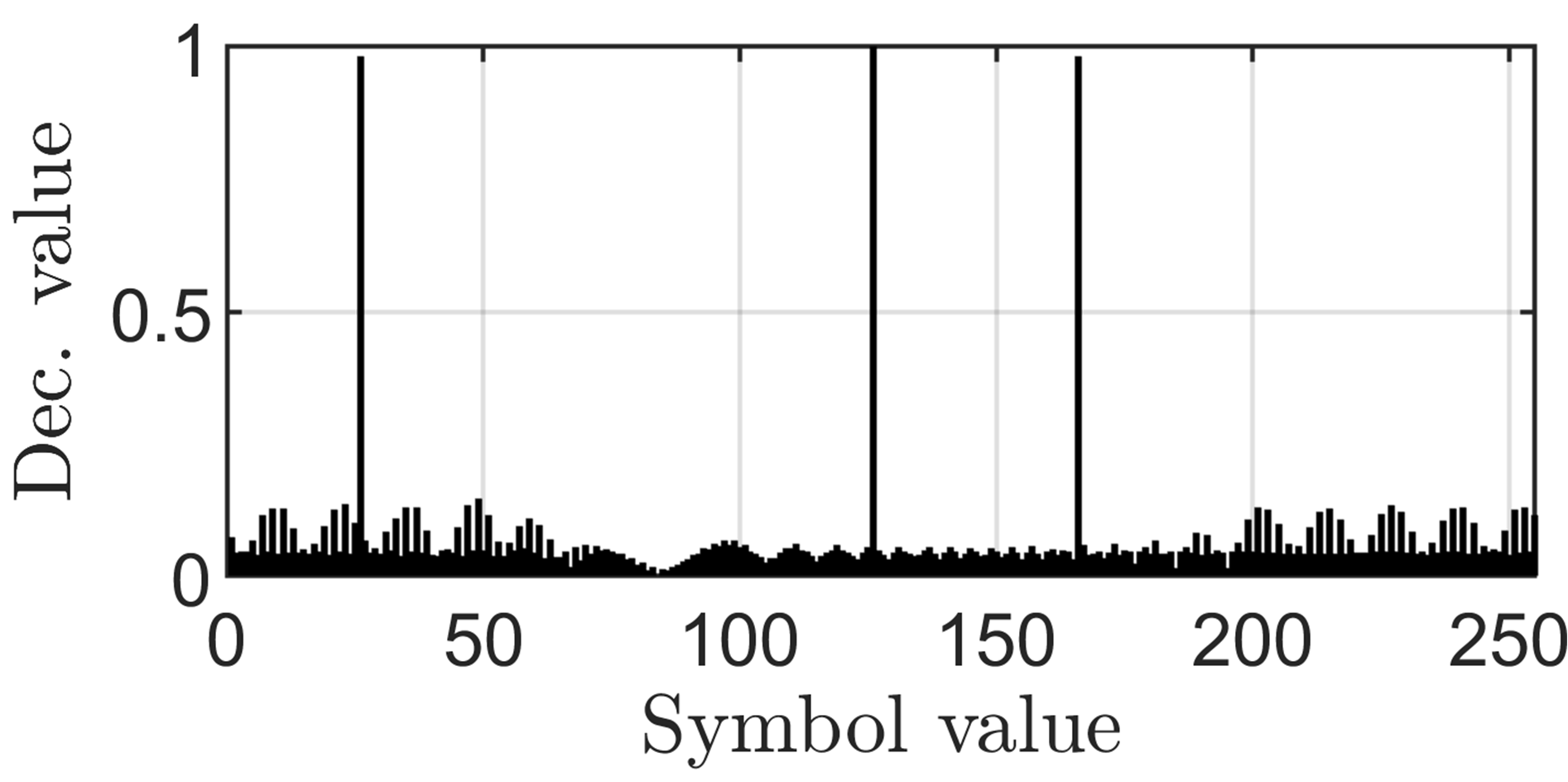}
		\caption{}
		\label{sub_fig:2}
	\end{subfigure}
	\caption{\justifying LoRa receiver decision value. (a) Separated desired signal and interference, where the desired signal is characterized by  $\{\text{encoded symbol},SF\}=\{25,8\}$, co-SF interference 1 by $\{125,8\}$, co-SF interference 2 by $\{165,8\}$, inter-SF interference 2 by $\{125,7\}$), inter-SF interference 2 by $\{165,9\}$). (b) Combined signal and interference.}
\end{figure}

The interference from EDs utilizing the same SF as the desired ED, termed co-SF interference, has power given by
\begin{equation}\label{I_0}
	I_1^{(m)}=\max\limits_{k^{(m)}\in\mathcal{P}_A^{(m)}\setminus \{0\}}\Upsilon^{(m)}\frac{P_\varsigma}{\big(r_{k^{(m)}}\big)^{\beta}}\left|h_{k^{(m)},l}\right|^2,
\end{equation}
where $\Upsilon^{(m)}$ is the co-SF interference sensitivity given in TABLE~\ref{tab:sir_thresholds} for $\tilde{m}=m$. The $\max\{\cdot\}$ operation stems from the inherent behavior of LoRa receiver for co-SF signals\footnote{Fig. \ref{sub_fig:1} demonstrates that the LoRa receiver resolves co-SF signals with different encoded symbols into distinct, non-overlapping peaks \cite{bibitem5}, \cite{bibitem11}. The desired signal is considered successfully recovered if the maximum of the co-SF interference remains below that of the desired signal.}.
\begin{table}[t]\label{TAB1}
	\centering
	\caption{INTERFERENCE THRESHOLDS WITH SX1272 TRANSCEIVER \\IN  DECIBEL (DB) \cite[TABLE II]{bibitem5}}
	\label{tab:sir_thresholds}
	\begin{tabular}{|c|c|c|c|c|c|c|}
		\hline
		\diagbox{$\tilde{m}$}{${m}$} & {7} & {8} & {9} & {10} & {11} & {12} \\
		\hline
		7 & 1 & -8 & -9 & -9 & -9 & -9 \\
		\hline
		8 & -11 & 1 & -11 & -12 & -13 & -13 \\
		\hline
		9 & -15 & -13 & 1 & -13 & -14 & -15 \\
		\hline
		10 & -19 & -18 & -17 & 1 & -17 & -18 \\
		\hline
		11 & -22 & -22 & -21 & -20 & 1 & -20 \\
		\hline
		12 & -25 & -25 & -25 & -24 & -23 & 1 \\
		\hline
	\end{tabular}
\end{table}
Moreover, the power of the inter-SF interference, where $\tilde{m}\neq{m}$, due to the imperfect orthogonality between SFs, is given by
\begin{equation}\label{I_1}
		\!\!I_2^{(m,\tilde{m})}=\!\!\sum\limits_{k^{(\tilde{m})}\in\mathcal{P}_A^{(\tilde{m})}}\!\!\!\Upsilon^{(m,\tilde{m})}\frac{P_\varsigma}{\big(r_{k^{(\tilde{m})}}\big)^{\beta}}\left|h_{k^{(\tilde{m})},l}\right|^2.
\end{equation}
Unlike co-SF interference, inter-SF interference exhibits fundamentally different mutual influence characteristics\footnote{The quasi-orthogonality between different SFs prevents the formation of a demodulation peak from an inter-SF signal (see Fig. \ref{sub_fig:1}). Instead, its power is spread across all symbol positions. Mathematically, this cumulative effect is modeled as a summation (see Fig. \ref{sub_fig:2}), leading some studies to approximate inter-SF interference as additive noise \cite{bibitem3}, \cite{bibitem5}, \cite{bibitem7}.}.

Another characteristic of LoRaWAN is the differential sensitivity $\Upsilon^{(m,\tilde{m})}$ among SF pairs, which arises from the different degrees of orthogonality between SFs. Specifically, the values of $\Upsilon^{(m,\tilde{m})}$ are provided in TABLE~\ref{tab:sir_thresholds}, which summarizes the experimental results for the LoRa SX1272 transceiver reported in \cite{bibitem5}.

Considering both the co-SF and inter-SF interference, the average coverage probability is expressed as
\begin{equation}\label{P_cov}
	\begin{split}
	&\overline{P}_\mathrm{cov}^{(m)}=\mathbb{P}\left\{\mathrm{SIR}_1\geq 1\cup \mathrm{SIR}_2\geq 1\cup \mathrm{SNR}\geq \Upsilon_{\mathrm{n}}^{(m)}\right\}\\
	&=\mathbb{P}\bigg\{S_0\geq\max\bigg\{I_1^{(m)},\sum\limits_{\tilde{m}=7,\tilde{m}\neq{m}}^{12}I_2^{(m,\tilde{m})},\Upsilon_{\mathrm{n}}^{(m)}\sigma^2\bigg\}\bigg\},
	\end{split}
\end{equation}
where $\mathrm{SIR}_1=\frac{S_0}{I_1^{(m)}}$, $\mathrm{SIR}_2=\frac{S_0}{\sum\limits_{\tilde{m}=7,\tilde{m}\neq{m}}^{12}I_2^{(m,\tilde{m})}}$ and $\mathrm{SNR}=\frac{S_0}{\sigma^2}$. Dropping the inter-SF interference term,  \eqref{P_cov} is accord with \cite[(3)]{bibitem9}. Besides, $\sigma^2=-174+NF+10\log_{10}(BW)$ in dBm is the variance of the additive white Gaussian noise (AWGN), and $NF$ is the noise figure of the receiver. Additionally, the network requires SNR to exceed the quality of service (QoS) threshold, which is given by $\Upsilon_{\mathrm{n}}^{(m)}=-7.5-2.5(m-7)$ dB\footnote{Increasing the SF in LoRa modulation enhances signal robustness against noise, improving reliability and enabling communication at higher noise levels.} \cite[Table 13]{bibitem17}. This implies that, owing to the physical-layer LoRa modulation and the correlation-based demodulation at the receiver, successful reception can still be achieved even when the QoS threshold is negative (in dB).

\section{Performance Analysis}\label{sec:analysis}
\subsection{Channel Approximation}\label{sec:h-approx}
\begin{figure}[!t]
	\centering
	\includegraphics[width=1\linewidth]{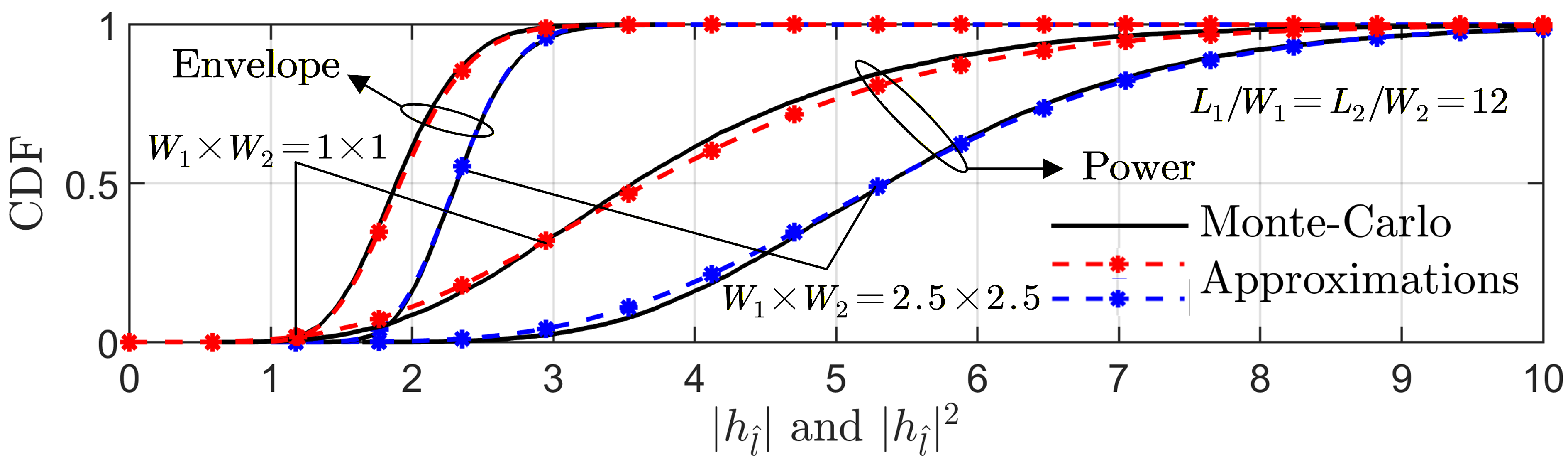}
	\caption{Approximations of envelope and power for the FAS channel.}\label{fig:3}
\end{figure}
Here, we initially derive a tractable approximation
for the FAS channel statistical distributions by employing the extreme value theory (EVT) and moment-matching method. Based on the results in \cite[Lemma 1]{bibitem4}, the CDF of maximum channel gain within a single block in \eqref{h-channel} can be expressed as 
\begin{equation}\label{h-approx}
		F_{\left|h_{0,(a,\hat{b}_a)}\right|}(r)\approx\left[ 1 - \exp\left(-\frac{(r-\delta_{a})^2}{\mu_a^2}\right) \right] \mathrm{U}(r-\delta_{a}).
\end{equation}
where $\big|h_{0,(a,\hat{b}_a)}\big|=\arg\max\limits_{b_a\in\{1,\cdots,L_a\}}\left|h_{0,(a,b_a)}\right|$ and $\mathrm{U}(\cdot)$ is the unit step function. For large $\mu_a^2$, \eqref{h-approx} can be regarded as a shift of the Rayleigh CDF, where the shift term is given by $\label{delta}\delta_{a}=\left({({1-\mu_a^2})/{2}\times\mathrm{W}\left({(L_a-2)^2}/{2\pi}\right)}\right)^{1/2}$ and $\mathrm{W}(\cdot)$ is the Lambert-W function.

Based on the above result, the CDF of multiple blocks, i.e., $\big|h_{0,\hat{l}}\big|$ can be approximated as the maximum of a set of Rayleigh RVs with different shift values, i.e., $\big|h_{0,\hat{l}}\big|=\max\big\{\big|h_{0,(1,\hat{b}_1)}\big|,\cdots,\big|h_{0,(A,\hat{b}_{A})}\big|\big\}$\footnote{In \cite{bibitem4}, an analytical expression for multiple blocks achieves satisfactory accuracy for linear FAS under Clarke’s model, where dominant eigenvalues are nearly equal. However, its accuracy degrades for Jake’s model or planar FAS under both Clarke’s and Jake’s models, and the VBCM, mainly due to unequal eigenvalues. Here, we propose an alternative approximation method.}. Utilizing the EVT for independent and non-identically distributed (i.n.i.d.) RVs \cite{bibitem18}, $\big|h_{0,\hat{l}}\big|$ can be approximated as $\mathrm{Gumbel}\left({\alpha_{A}},\beta_A\right)$, where $\alpha_{A}$ and $\beta_{A}$ are normalized factors chosen as
\begin{equation}\label{alpha_beta}
	\left\{
	\begin{split}
		\alpha_{A} &= \overline{\mu} F^{-1}_{\left|h_{0,\left(a,b_a\right)}\right|}\bigg(1-\frac{1}{A}\bigg)+\frac{1}{A}\sum_{a=1}^{A}\delta_{a}, \\
		\beta_{A} &= \overline{\mu}\left[F^{-1}_{\left|h_{0,(a,b_a)}\right|}\left(1-\frac{1}{A e}\right)-F^{-1}_{\left|h_{0,\left(a,b_a\right)}\right|}\left(1-\frac{1}{A}\right)\right],
	\end{split}
	\right.
\end{equation}
where $\overline{\mu}$ is the average of $\mu_a$ and $F^{-1}_{\left|h_{0,(a,b_a)}\right|}(\cdot)$ is the inverse CDF of $\mathrm{Rayleigh}\left(\frac{1}{\sqrt{2}}\right)$.
The proof is given in Appendix \ref{app-a}.

It is noted that the statistical expression of the Gumbel distribution is nearly analytically intractable. However, its higher-order moments can be readily obtained as 
\begin{equation}\label{moment-h}
	\begin{split}
		\mathbb{E}\big\{&\big|h_{0,\hat{l}}\big|^n\big\}=\frac{\partial^n \left\{\Gamma\left(1-{\beta}_{A} t\right)e^{{\alpha}_{A}t}\right\}}{\partial t^n}\Big|_{t=0}\\
		&\overset{(a)}{=}\sum_{\tilde{n}=0}^n\binom{n}{\tilde{n}}\alpha_{A}^{n-\tilde{n}}(-\beta)^{\tilde{n}}\mathrm{B}_{\tilde{n}}\left(\begin{aligned}&\psi(1),\psi^{(1)}(1),\\
		&\cdots,\psi^{({\tilde{n}}-1)}(1)\end{aligned}\right),
	\end{split}
\end{equation}
where $(a)$ is based on Leibniz's rule. Besides, $\psi^{(\cdot)}(\cdot)$ is the Polygamma function and $B_n(\cdot)$ denotes the exponential Bell polynomials, which can be directly obtained from the generating function or lookup table. 

Utilizing the moment-matching method, the distribution of $\big|h_{0,\hat{l}}\big|$ and $\big|h_{0,\hat{l}}\big|^2$ can be accurately approximated by a Gamma distribution, given by
\begin{equation}\label{h-cdf}
	\begin{split}
		\big|h_{0,\hat{l}}\big|^n\sim\mathrm{Gamma}(\theta_{0},\omega_{0}),
	\end{split}
\end{equation}
where $n\in\{1,2\}$. Moreover, the shape and scale parameters are defined as $\theta_{0}=\frac{\varphi_{0}^2}{\vartheta_{0}}$ and $\omega_{0}=\frac{\varphi_{0}}{\vartheta_{0}}$, where $\varphi_{0}=\mathbb{E}\big\{\big|h_{0,\hat{l}}\big|^n\big\}$ and $\varphi_{0}=\mathbb{E}\big\{\big|h_{0,\hat{l}}\big|^{2n}\big\}-\mathbb{E}\big\{\big|h_{0,\hat{l}}\big|^n\big\}^2$. Note that the moment  $\mathbb{E}\big\{\big|h_{0
	,\hat{l}}\big|^{n}\big\}$ is given by \eqref{moment-h}. 

Fig.~\ref{fig:3} presents the simulation results for Section~\ref{sec:h-approx}. Note that the eigenvalue-based EVT technique is generally more effective for larger FAS apertures. However, in practice, the proposed approximation still provides satisfactory accuracy for both relatively small and large normalized apertures, such as $W_1\times W_2=1\times1$ and $2.5\times2.5$.
\subsection{Coverage Probability Analysis}
In this subsection, the coverage probability of the proposed network is analyzed. According to  \eqref{P_cov}, it can be expressed as
\begin{equation}\label{int_P_cov}
	\overline{P}_\mathrm{cov}^{(m)}=\int_{\Upsilon_{\mathrm{n}}^{(m)}\sigma^2}^{\infty}F_{I_1^{(m)}}(x)F_{\tilde{I}_2^{(m)}}(x)f_{S_0}(x)dx,
\end{equation}
where $\tilde{I}_2^{(m)}=\sum\limits_{\tilde{m}=7,\tilde{m}\neq{m}}^{12}I_2^{(m,\tilde{m})}$ and the CDF of the maximum co-SF interference is given by \cite[Lemma 1]{bibitem9}
\begin{equation}\label{cdf-I1}
	\begin{split}
		\!\!\!&F_{I_1^{(m)}}(x)=F_{I_1^{(m)}}(x;R_2)-F_{I_1^{(m)}}(x;R_1)
		\!\\
		&\!\!\!=\!\exp\!\Bigg[\!\!-\!\frac{2\overline{N}^{(m)}}{\beta\left(R_2^2-R_1^2\right)}
		\left(\!\frac{P_\varsigma\Upsilon^{(m)}}{x}\!\right)^{\frac{2}{\beta}}
		\!\!\!\!\!\gamma\left(\frac{2}{\beta},\frac{x R_2^\beta}{P_\varsigma\Upsilon^{(m)}},\frac{x R_1^\beta}{P_\varsigma\Upsilon^{(m)}}\!\right)\!
		\Bigg],
	\end{split}
\end{equation}
where $\gamma(\cdot,b,a)=\gamma(\cdot,b)-\gamma(\cdot,a)$ and $\gamma(\cdot,\cdot)$ is the lower incomplete gamma function.

Next, the CDF of the sum of inter-SF interference can be obtained by applying the Laplace transform (LT) and its inverse, which is obtained by
	  $F_{\tilde{I}_2^{(m)}}(x)=\mathcal{L}^{-1}\left\{\mathcal{L}_{\tilde{I}_2^{(m)}}(s)/s\right\}(x)$, where the LT of $\tilde{I}_2^{(m)}$ is given by \cite[Appendix A]{bibitem12}
\begin{multline}
	\mathcal{L}_{\tilde{I}_2^{(m)}}(s)=\prod
	\limits_{\tilde{m}\in\mathcal{M},\tilde{m}\neq{m}}\\
	\exp\Bigg[-2\pi\lambda_A^{(\tilde{m})}\int_{R_1}^{R_2}\bigg(1-\frac{r^\beta}{r^\beta+s\Upsilon^{(m,\tilde{m})}P_\varsigma}\bigg)rdr\Bigg].
\end{multline}
However, an analytical expression for the inverse LT is not available. To facilitate the analysis, the fractional term in the integrand is expanded into an infinite series as
\begin{equation}\label{series}
	\frac{{r}^{\beta}}{{r}^{\beta}+s\Upsilon^{(m,\tilde{m})}P_\varsigma}=\sum\limits_{n=0}^{\infty}(-1)^n\Big(s\Upsilon^{(m,\tilde{m})}P_\varsigma{r}^{-\beta}\Big)^n.
\end{equation}
By retaining the first two terms in \eqref{series} and applying the inverse LT, the CDF of $\tilde{I}_2^{(m)}$ can be approximated as
\begin{equation}\label{cdf-I2}
		F_{\tilde{I}_2^{(m)}}({x})\approx
		\mathcal{L}^{-1}\left\{{\exp\left(-\zeta^{({m})}s\right)}/{s}\right\}(x)=\mathrm{U}\left(x-\zeta^{({m})}\right),
\end{equation}
where $\zeta^{({m})}=
\sum\limits_{\tilde{m}\in\mathcal{M},\tilde{m}\neq m}{2\pi\lambda_A^{(\tilde{m})}\Upsilon^{(m,\tilde{m})} P_\varsigma}\frac{R_2^{2-\beta}\!-\!R_1^{2-\beta}}{2-\beta}
$. It is worth noting that $\zeta^{(m)}$ equals the mean of $\tilde{I}_2^{(m)}$, i.e.,
\begin{equation}
	\begin{split}\label{mean_I2}
		&\mathbb{E}\left\{\tilde{I}_2^{(m)}\right\}\\
		&=\sum\limits_{\tilde{m}\in\mathcal{M},\tilde{m}\neq{m}}\mathbb{E}\Bigg\{\sum\limits_{k^{(\tilde{m})}\in\mathcal{P}_A^{(\tilde{m})}}\Upsilon^{(m,\tilde{m})}\frac{P_\varsigma}{\big(r_{k^{(\tilde{m})}}\big)^{\beta}}\left|h_{k^{(\tilde{m})},l}\right|^2\Bigg\}\\
		&=\sum\limits_{\tilde{m}\in\mathcal{M},\tilde{m}\neq{m}}\lambda_A^{(\tilde{m})}\Upsilon_2^{(m,\tilde{m})}P_\varsigma\int_0^{2\pi}\int_{R_1}^{R_2}r^{1-\beta}drd\theta\equiv\zeta^{(m)}.
	\end{split}
\end{equation}
This approximation is justified because the quasi-orthogonality among SFs yields small weights $\Upsilon^{(m,\tilde{m})}$ of ${I}_2^{(m,\tilde{m})}$ as indicated in TABLE \ref{tab:sir_thresholds}. Therefore, the variation of $\tilde{I}_2^{(m)}$ is much smaller than that of $I_1^{(m)}$ and $S_0$, and thus $\tilde{I}_2^{(m)}$ can be treated as a constant. It is also consistent with Chebyshev’s law of large numbers in \cite[Lemma 3]{bibitem19}.

Next, the PDF of $S_0$ is given by
\begin{equation}\label{pdf-s}
\begin{split}
&f_{S_0}(x)\\&=\frac{\partial}{\partial x}\left\{\frac{2}{\beta(R_2^2-R_1^2)}\int_{R_1^\beta}^{R_2^\beta}\bar{\gamma}\left(\theta_0,\frac{rx}{P_\varsigma\omega_0}\right)r^{\frac{2}{\beta}-1}dr\right\}\\
&=\xi(x;R_2)-\xi(x;R_1),
\end{split}
\end{equation}
where $\overline{\gamma}(a,b)={{\gamma}(a,b)}/{\Gamma(a)}$, $\Gamma(\cdot)$ is the Gamma function. Let $\chi\in\{1,2\}$, we have
\begin{equation}\label{xi}
		\xi(x;R_{\chi})
		= x^{-\frac{2}{\beta}-1}\, C\,\overline{\gamma}\left(\varrho,\varphi(R_{\chi})x\right),
\end{equation}
where $C= \frac{2\left(\theta_0\right)_{\frac{2}{\beta}}}{\beta\left(R_2^2-R_1^2\right)}(P_\varsigma\omega_0)^{\frac{2}{\beta}}$ and $(a)_{b}$ is the Pochhammer symbol. Besides, $\varrho = \theta_0+\frac{2}{\beta}$ and $\varphi(R_\chi)={R_\chi^{\beta}}/{P_\varsigma\omega_0}$. 
Then, using the linear interpolation and finite series expansion on the $\overline{\gamma}(\cdot,\cdot)$ term in \eqref{xi}, we have
\begin{equation}\label{gamma_approx}
	\overline{\gamma}\left(\varrho,\varphi\left(R_{\chi}\right)x\right)\approx1-e^{-\varphi(R_\chi)x}\sum\limits_{i=0}^{\lfloor\varrho\rfloor}c_i\frac{[\varphi(R_{\chi})x]^i}{i!},
\end{equation}
where $
	c_i = 
	\begin{cases}
		1, & 0 \leq i < \lfloor\varrho\rfloor \\
		\varrho - \lfloor\varrho\rfloor, & i = \lfloor\varrho\rfloor
	\end{cases}
	$ and $\lfloor \cdot \rfloor$ denotes the floor operation.

In \eqref{cdf-I1}, the CDF of $I_2^{(\tilde{m})}$ is lower bounded by taking the maximum of the difference of gamma functions, as
\begin{equation}\label{approx-cdf-I1}
	\begin{split}
		&F_{I_1^{(m)}}(x)\\
		&\!\approx\exp\Bigg[\!-\!\underbrace{\frac{2\overline{N}_m(P_\varsigma\Upsilon^{(m)})^{\frac{2}{\beta}}}{\beta\left(R_1^2-R_2^2\right)}\gamma\left(\frac{2}{\beta},\frac{\hat{x} R_2^\beta}{P_\varsigma\Upsilon^{(m)}},\frac{\hat{x} R_1^\beta}{P_\varsigma\Upsilon^{(m)}}\right) }_{D^{(m)}}
		x^{-\frac{2}{\beta}}
		\Bigg],
	\end{split}
\end{equation}
where $\hat{x}=\frac{2P_\varsigma\ln(R_2/R_1)}{R_2^\beta-R_1^\beta}$. By combining \eqref{int_P_cov}, \eqref{pdf-s}, \eqref{xi}, \eqref{gamma_approx} and \eqref{approx-cdf-I1}, the coverage probability can be expressed as
\begin{equation}
	\overline{P}_\mathrm{cov}^{(m)} = \varTheta^{(m)}(R_1) - \varTheta^{(m)}(R_2),
\end{equation}
where $\varTheta^{(m)}(R_\chi)=CL_1^{(m)}
	-C\sum\limits_{i=0}^{\lfloor\varrho\rfloor}c_i\frac{[\varphi(R_{\chi})]^i}{i!}L^{(m)}_{2,i}(R_\chi)$.
Then, the final two integrals are given by
\begin{multline}\label{int1}
	\color{black}L_1^{(m)}=\int_{\tilde{\zeta}^{({m})}}^{\infty}\exp\left({-D^{(m)}x^{-\frac{2}{\beta}}}\right)x^{-\frac{2}{\beta}-1}dx\\
	\color{black}=\frac{\beta}{2D^{(m)}}\left\{1-\exp\left[-D^{(m)}\left(\tilde{\zeta}^{({m})}\right)^{-\frac{2}{\beta}}\right]\right\},
\end{multline}
\textcolor{black}{where $\tilde{\zeta}^{(m)}=\max(\zeta^{(m)},\Upsilon_{\mathrm{n}}^{(m)}\sigma^2)$. \eqref{int1} is obtained by a variable substitution. Moreover, $L_2$ can be expressed as}
\begin{multline}\label{L_2}
		\!\!\!L_{2,i}^{(m)}(R_\chi)=\! \int_{\tilde{\zeta}^{({m})}}^{\infty}\!\!\!\exp\left(-D^{(m)} x^{-\frac{2}{\beta}}\!-\!\varphi\left(R_\chi\right) x \right) x^{-\frac{2}{\beta} - 1 + i}dx \\
		=\frac{\beta}{2}\left\{{ \varphi\left(R_\chi\right)^{-\nu}} \mathrm{H}^{2,0}_{0,2}\left[
		\varphi\left(R_\chi\right) (D^{(m)})^{\frac{\beta}{2}} \,\middle|\,
		\begin{array}{c}
			- \\
			\left(0, \tfrac{\beta}{2}\right), \left(\nu, 1\right)
		\end{array}
		\right]\right.\\
		-\sum_{k=0}^\infty\frac{\left[-\varphi\left(R_\chi\right)\right]^k}{k!} (D^{(m)})^{\frac{\beta(k+i)}{2}-1}\\
		\left.
		\times\Gamma\left[1-\frac{\beta(k+i)}{2},\frac{D^{(m)}}{\big(\tilde{\zeta}^{({m})}\big)^{\frac{2}{\beta}}}\right]\right\},
\end{multline}
where $\nu=i-\frac{2}{\beta}$. $\mathrm{H}^{m,n}_{p,q}\left[\cdot \right]$ is the Fox H-function, obtained by transforming the integral limits and using \cite[(6.23)]{bibitem20}. Moreover, $\Gamma(\cdot,\cdot)$ is the upper  Gamma function, obtained by \cite[3.3811.3]{bibitem21}.

For the special case without inter-SF interference and with large SF, i.e., $\tilde{\zeta}^{(\tilde{m})}=0$, the infinite-series term in \eqref{L_2} vanishes, and a closed-form solution can be derived.
\section{Numerical Results}\label{sec:results}	
\begin{figure}[!t]
	\centering
	\begin{subfigure}[b]{0.49\textwidth}
		\centering
		\includegraphics[width=1\textwidth]{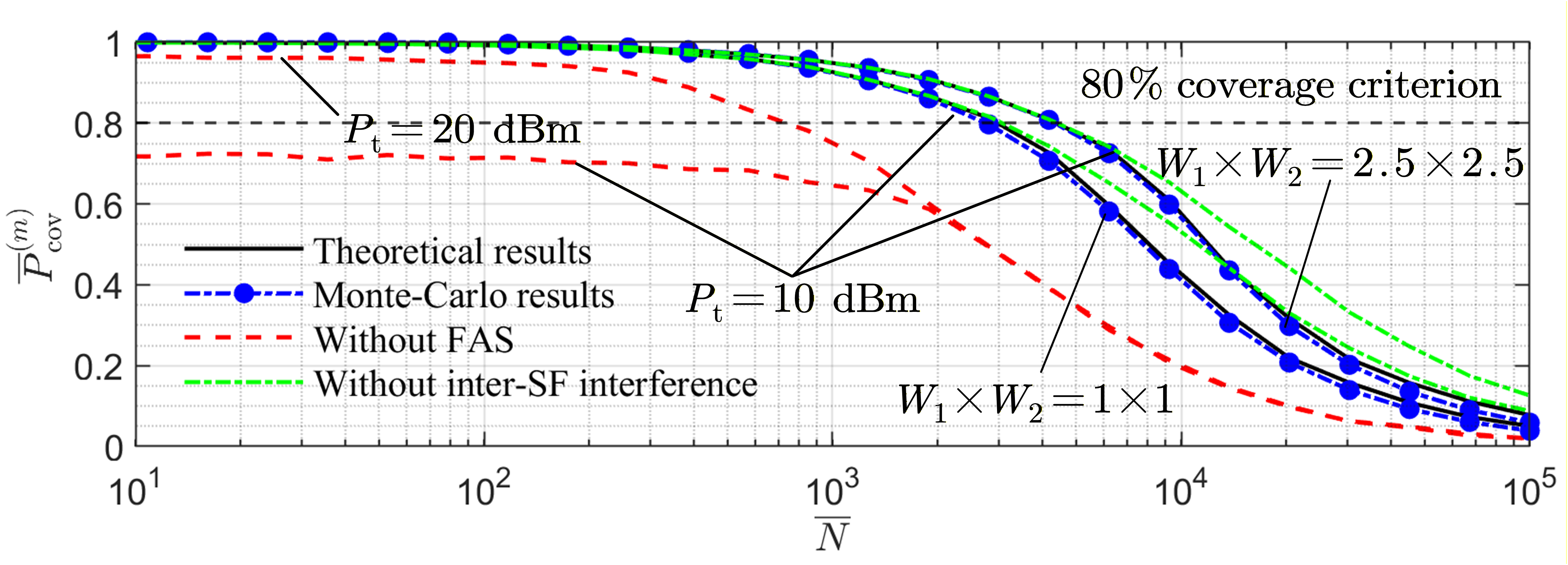}
		\caption{}
		\label{sub_fig:4}
	\end{subfigure}
	\vspace{2pt}
	\begin{subfigure}[b]{0.49\textwidth}
		\centering
		\includegraphics[width=1\textwidth]{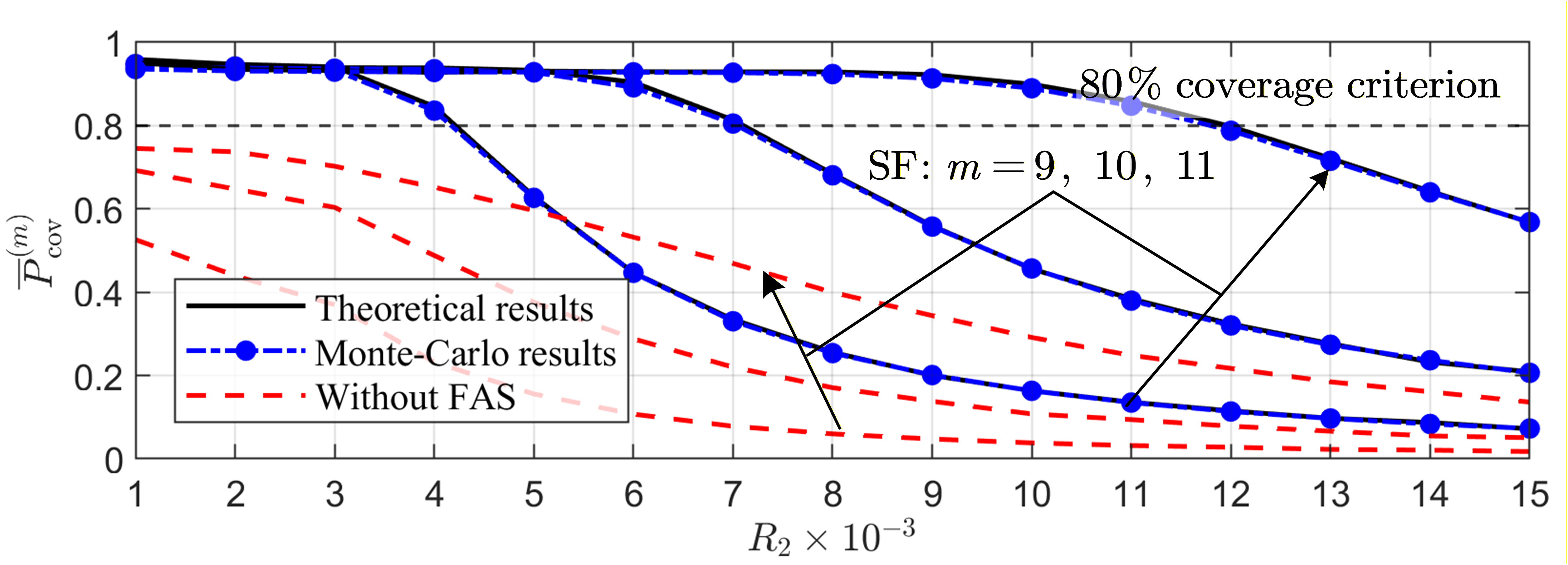}
		\caption{}
		\label{sub_fig:5}
	\end{subfigure}
	\vspace{-14pt}
	\caption{\justifying Coverage probability of the FAS-enhanced LoRa network under 
		(a) varying numbers of EDs ($m=9$, $P_\mathrm{t}=\{10, 20\}~\mathrm{dBm}$, $W_1\times W_2=\{1\times1, 2.5\times2.5\}$, $R_2=3000~\mathrm{m}$), and 
		(b) varying network region radius ($m=\{9, 10, 11\}$, $P_\mathrm{t}=10~\mathrm{dBm}$, $W_1\times W_2=1.5\times1.5$, $\overline{N}=10^3$).}
\end{figure}
In this section, we evaluate the performance of the proposed network via Monte Carlo simulations and theoretical values. Unless otherwise specified, the simulation parameters are set to  $L_\mathrm{pl}=20~\mathrm{bytes}$, $N_\mathrm{pre}=150$, $L_\mathrm{hd}=20$, $L_\mathrm{crc}=16$, $CR=1$, $T_\text{in}=400$, $f_\mathrm{c}=915~\mathrm{MHz}$ ($\lambda\approx32.8~\mathrm{cm}$), $BW=125~\mathrm{kHz}$, $\beta=2.2$, $NF=6~\mathrm{dB}$ $R_1=300~\mathrm{m}$ and $L_1/W_1=L_2/W_2=12$. Additional parameters are provided in the corresponding figure captions. Each result is obtained from $10^4$ Monte Carlo realizations.

Fig.~\ref{sub_fig:4} depicts the coverage performance of the proposed network versus the number of EDs. First, the analytical results closely match the Monte Carlo simulations, validating the rationality of the approximation of the co-SF interference. Second, a FAS with a compact aperture, such as $W_1\times W_2=1\times1$, demonstrates substantial performance gains over the single-antenna scheme, outperforming even the $10~\mathrm{dB}$ power-enhanced case. Besides, due to the impact of interference, the benefit of power enhancement alone diminishes as the number of EDs increases. However, the FAS can realize passive interference suppression by increasing only the desired signal power. Third, the improvement from no FAS to a FAS with the normalized size of $1\times1$ is significantly larger than that achieved by further increasing the normalized size from $1\times1$ to $2.5\times2.5$.

In Fig.~\ref{sub_fig:5}, the coverage probability gain increases with SF, which is consistent with LoRa physical characteristics, since a larger SF yields more sample points, enabling greater accumulated gain from the FAS through correlation detection at the gateway. Moreover, in conventional LoRa networks, increasing the SF is the primary means of extending coverage range. By contrast, FAS provides a more pronounced coverage improvement than increasing the SF alone, highlighting its potential in LoRa networks. It should be noted that the above results are mainly presented as a performance reference for FAS in LoRa networks. In practical deployments, the achievable performance may be limited by channel estimation and port-switching overhead. The impact of these overheads further depends on the channel coherence time. Nevertheless, unlike mobile scenarios, LoRa EDs are typically quasi-static and deployed at fixed locations, which helps reduce the overhead required for the operation of FAS.
\section{Conclusion}\label{sec:conclude}
 This paper has developed an analytical framework to evaluate the coverage performance of FAS-enhanced LoRaWAN. We considered the effects of both large-scale pathloss and small-scale fading for the FAS channel. The impact of co-SF and inter-SF interference of LoRaWAN is also considered. A simple yet efficient approximate expression for FAS is proposed, and based on the approximation, we derive the analytical coverage probability for FAS-enhanced LoRaWAN. The numerical results demonstrate that, even in the presence of density interference, a FAS with a small normalized aperture, such as $1\times1$, can still significantly enhance network coverage performance by exploiting spatial potential.
\appendices
\section{}\label{app-a}
Before the proof, we first revisit the EVT, which states that the maximum of a sequence of i.n.i.d. RVs $\{H_1, \cdots, H_A\}$ converges weakly to a non-degenerate distribution function with the CDF of ${F}(x)=e^{u(x)}$, where
\begin{equation}\label{UA2}
	\begin{split}	
		u(x)=\lim\limits_{A\to\infty}\sum_{a=1}^{\infty}1-F_{H_A}\left(\alpha_Ax+\beta_A\right)<\infty,
	\end{split}
\end{equation}	
if the sequences and normalizing factors $\{\alpha_1,\cdots,\alpha_A\}$ and $\{\beta_1,\cdots,\beta_A\}$ satisfies the Meizler's condition \cite{bibitem22}, which is given by (i) $C_1=\lim\limits_{{A}\to\infty}\big|\log\left(\alpha_{A}\right)\big|+\big|\beta_{N}\big|\to\infty$; (ii) $C_2=\lim\limits_{{A}\to\infty}\alpha_{A+1}\!-\!\alpha_{A}\!\to\!1$; (iii) $C_3\!=\!\lim\limits_{{A}\!\to\infty}({\beta_{A+1}-\beta_A})/\alpha_A\!\to\!0$.

This appendix aims to prove the conditions satisfied and calculating the $u(x)$ in \eqref{UA2} to get the extreme distribution. Let $H_a=H_{\mathrm{Ra},a}+\delta_{a}$, for $a=\{1,\cdots,A\}$. Moreover, let $H_{\mathrm{Ra},a}\sim\mathrm{Rayleigh}(1/\sqrt{2})$. Then the inverse CDF of ${H}_{\mathrm{Ra},a}$ is given by $F^{-1}_{\left|H_{\mathrm{Ra},a}\right|}(x)={-2\mathrm{log}(1-x)}^{1/2}$. Then $\tilde{H}_A=\max\{{H}_1,\cdots,{H}_A\}$ has a similar CDF of \eqref{h-approx}. Next, let $\alpha_A$ and $\beta_A$ are chosen as \eqref{alpha_beta}, we have
		$C_1=\big|\log\left(\alpha_{A}\right)\big|+\big|\beta_{A}\big|\geq\mu\sqrt{A}\to\infty$,
and
		$C_2=\lim\limits_{{A}\to\infty}\frac{\sqrt{\log\left({A}+1\right)+1}-\sqrt{\log\left({A}+1\right)}}{\sqrt{\log\left({A}\right)+1}-\sqrt{\log\left({A}\right)}}=1$,
which is obtained by using the limits law. Moreover, we have
\begin{equation}\label{UA1-3}
	\begin{split}	
	C_3&=\!\lim\limits_{A\to\infty}\!
		\frac{\!\!\!\begin{array}{lcl}
				\sqrt{\log\left({A}+1\right)}-\!\sqrt{\log\left({A}\right)}+\Delta(A,A+1)
			\end{array}}
		{\sqrt{\log\left({A}\right)+1}-\sqrt{\log\left({A}\right)}}\\
		&=\lim\limits_{A\to\infty}
		\frac{\begin{array}{lcl}
				\Delta(A,A+1)
		\end{array}}
		{\sqrt{\log\left({A}\right)+1}-\sqrt{\log\left({A}\right)}},
	\end{split}
\end{equation}	
where $\Delta(A,A+1)=\frac{1}{A+1}\sum_{a=1}^{A+1}\delta_{a}-\frac{1}{A}\sum_{a=1}^{A}\delta_{a}$. Here, we assume that $W_1/L_1=W_2/L_2=\xi$ is a large constant. Then, $A$ will increase as $W_1$ and $W_2$. Moreover, we have $\sum_{a=1}^A L_a\approx\xi^2W_1W_2=\xi^2W$. This is because these eigenvalues dominate the correlation matrix. As $\delta_{a}$ is a function of $L_a$, it can be prove that $\sum_{a=1}^A\delta_{a}$ is a linear function of $W$. Besides, $A$ can also be approximated as a linear function of $W$ \cite{bibitem23}. Then we assume that $\sum_{a=1}^A\delta_{L_a}/A=(xA+y)/A$, where $x$ and $y$ is the parameter. Then \eqref{UA1-3} can be reformulated as
\begin{equation}\label{UA1-3-1}
	\begin{split}	
		C_3=\lim\limits_{A\to\infty}
		\frac{\begin{array}{lcl}
			\frac{x(A+1)+y}{A+1}-\frac{xA+y}{A}
		\end{array}}
		{\sqrt{\log\left({A}\right)+1}-\sqrt{\log\left({A}\right)}}=0,
	\end{split}
\end{equation}	
and we also have $\lim\limits_{A\to\infty}\Delta(A,A+1)=0$. 

Finally, substitute \eqref{alpha_beta} into \eqref{UA2}, $F_{\tilde{H}_A}(x)=e^{-e^{-x}}$ is accord with the Gumbel distribution.
\bibliographystyle{IEEEtran}

\begin{thebibliography}{10}
	\providecommand{\url}[1]{#1}
	\csname url@samestyle\endcsname
	\providecommand{\newblock}{\relax}
	\providecommand{\bibinfo}[2]{#2}
	\providecommand{\BIBentrySTDinterwordspacing}{\spaceskip=0pt\relax}
	\providecommand{\BIBentryALTinterwordstretchfactor}{4}
	\providecommand{\BIBentryALTinterwordspacing}{\spaceskip=\fontdimen2\font plus
		\BIBentryALTinterwordstretchfactor\fontdimen3\font minus
		\fontdimen4\font\relax}
	\providecommand{\BIBforeignlanguage}[2]{{%
			\expandafter\ifx\csname l@#1\endcsname\relax
			\typeout{** WARNING: IEEEtran.bst: No hyphenation pattern has been}%
			\typeout{** loaded for the language `#1'. Using the pattern for}%
			\typeout{** the default language instead.}%
			\else
			\language=\csname l@#1\endcsname
			\fi
			#2}}
	\providecommand{\BIBdecl}{\relax}
	\BIBdecl
	
	\bibitem{bibitem1}
	L.~Chettri and R.~Bera, ``A comprehensive survey on internet of things (IoT) toward 5G wireless systems,'' \emph{IEEE Internet Things J.}, vol.~7, no.~1, pp. 16--32, Jan. 2020.
	\bibitem{bibitem2}
	T.~Elshabrawy and J.~Robert, ``Closed-form approximation of LoRa modulation BER performance,'' \emph{IEEE Commun. Lett.}, vol.~22, no.~9, pp. 1778--1781, Sep. 2018.
	\bibitem{bibitem3}
	Z. Liang \textit{et al.}, ``RIS-Enabled Anti-Interference in LoRa Systems,'' \emph{IEEE Trans. Commun.}, vol. 72, no. 10, pp. 6599-6616, Oct. 2024.
	\bibitem{bibitem4}
	G.~Mu \textit{et al.}, ``On Performance of LoRa Fluid Antenna Systems,'' \emph{ IEEE Trans. Wireless Commun.}, \url{DOI: 10.1109/TWC.2025.3614671}, 2025.
	\bibitem{bibitem5}
	D. Croce \textit{et al.}, ``Impact of LoRa Imperfect Orthogonality: Analysis of Link-Level Performance,'' \emph{ IEEE Commun. Lett.}, vol. 22, no. 4, pp. 796-799, Apr. 2018.
	\bibitem{bibitem6}
	Q. M. Qadir \textit{et al.}, ``Analysis of the Reliability of LoRa,''  \emph{ IEEE Commun. Lett.}, vol. 25, no. 3, pp. 1037-1040, Mar. 2021.
	\bibitem{bibitem7}
	O. Afisiadis \textit{et al.}, ``On the Advantage of Coherent LoRa Detection in the Presence of Interference,'' \emph{IEEE Int. Things J.}, vol. 8, no. 14, pp. 11581-11593, Jul. 2021.
	\bibitem{bibitem8}
	A. Mahmood \textit{et al.}, ``Scalability Analysis of a LoRa Network Under Imperfect Orthogonality,'' \emph{IEEE Transactions Ind. Inf.}, vol. 15, no. 3, pp. 1425-1436, Mar. 2019.
	\bibitem{bibitem9}
	L.~Tu \textit{et al.}, ``On the Spectral Efficiency of LoRa Networks: Performance Analysis, Trends and Optimal Points of Operation,'' \emph{IEEE Trans. Commun.}, vol. 70, no. 4, pp. 2788-2804, Apr. 2022.
	\bibitem{bibitem10}
	Q. Yu \textit{et al.}, ``Toward LoRa-Based LEO Satellite IoT: A Stochastic Geometry Perspective,'' \emph{IEEE Internet Things J.}, vol. 12, no. 15, pp. 30725-30738, Aug. 2025.
	\bibitem{bibitem11}
	L. Beltramelli \textit{et al.}, ``LoRa Beyond ALOHA: An Investigation of Alternative Random Access Protocols,'' \emph{IEEE Trans. Ind. Inf.}, vol. 17, no. 5, pp. 3544-3554, May 2021.
	\bibitem{bibitem12}
	Q. Cheng \textit{et al.}, ``Design and Performance Analysis of MEC-Aided LoRa Networks With Power Control,'' \emph{IEEE Trans. Veh. Technol.}, vol. 74, no. 1, pp. 1597-1609, Jan. 2025.
	\bibitem{bibitem13}
	Z. Qin \textit{et al.}, ``Performance Analysis of Clustered LoRa Networks,'' \emph{IEEE Trans. Veh. Technol.}, vol. 68, no. 8, pp. 7616-7629, Aug. 2019.
	\bibitem{bibitem14}
	B. Reynders \textit{et al.}, ``Power and spreading factor control in low power wide area networks,'' in \emph{Proc. IEEE Int. Conf. Commun. (ICC)}, Paris, France, May, 2017, pp. 1-6.
	\bibitem{bibitem15}
	P.~Ram\'{i}rez-Espinosa \textit{et al.}, ``A new spatial block-correlation model for fluid antenna systems,''  \emph{ IEEE Trans. Wireless Commun.}, \url{DOI: 10.1109/TWC.2024.3434509}, 2024.
	\bibitem{bibitem16}
	X. Lai \textit{et al.}, ``Revisiting Spatial Block-Correlation Model for Fluid Antenna Systems: from Constant to Variable Correlations,'' \emph{ IEEE J. sel. Areas Commun.}, \url{DOI: 10.1109/JSAC.2025.3617021}, 2025.
	\bibitem{bibitem17}
	{\it LoRa SX1272/73 Transceiver Datasheet} , Semtech Co., Carmarillo, CA, USA, 2015. \href{https://semtech.my.salesforce.com/sfc/p/#E0000000JelG/a/440000001NCE/v_VBhk1IolDgxwwnOpcS_vTFxPfSEPQbuneK3mWsXlU}{[Online]. Available.}
	\bibitem{bibitem18}
	H. M. Barakat \textit{et al.}, ``Limit theorems for random maximum of independent and non-identically distributed random vectors,'' \emph{Statistics}, vol. 47, no. 3, pp. 546-557, 2013.
	\bibitem{bibitem19}
	Z. Dai \textit{et al.}, ``Coverage Probability and Average Rate Analysis of Hybrid Cellular and Cell-free Network,'' \emph{ IEEE Trans. Wireless Commun.}, \url{DOI: 10.1109/TWC.2025.3616960}, 2025.
	\bibitem{bibitem20}
	A.M. Mathai \textit{et al.}, ``The H-Function, Theory and Applications,''Springer Science \& Business Media, 2009.
	\bibitem{bibitem21}
	I.~S.~Gradshteyn and I.~M.~Ryzhik. ``Table of integrals, series, and products'', Academic press, 2014.
	\bibitem{bibitem22}
	C. Forbes, et. al., ``Statistical distributions,'' John Wiley \& Sons, 2011.
	\bibitem{bibitem23}
	E. Björnson and L. Sanguinetti, ``Rayleigh Fading Modeling and Channel Hardening for Reconfigurable Intelligent Surfaces,'' \emph{IEEE Wireless Commun. Lett.}, vol. 10, no. 4, pp. 830-834, Apr. 2021.
\end{thebibliography}

\end{document}